\documentclass[aps,amsmath,showpacs,amsfonts,10pt]{revtex4}
\usepackage{epsfig,graphicx}
\usepackage[english]{babel}
\usepackage{amsfonts}
\usepackage{amsmath}
\usepackage{latexsym}
\usepackage{graphics,bm}
\usepackage{natbib}
\usepackage{dcolumn}
\usepackage{bm}
\usepackage{rotating}

\begin{document}

\title{Damping in two component Bose gas}

\author{Aranya B Bhattacherjee$^{1,2}$}

\address{$^{1}$School of Physical Sciences, Jawaharlal Nehru University, New Delhi-110067, India} \address{$^{2}$Department of Physics, ARSD College, University of Delhi (South Campus), New Delhi-110021, India}

\begin{abstract}
We investigate the Landau and Baliaev damping of the collective modes in a two-component Bose gas using the mean-field approximation. We show that due to the two body atom-atom interaction, oscillations of each component is coupled to the thermal excitations of the other component which gives rise to creation or destruction of the elementary excitations that can take place in the two separate components.In addition we find that the damping is also enhanced due to inter-component coupling.

\end{abstract}


\maketitle

\section{Introduction}
The elementary excitations of trapped Bose-Einstein condensate (BEC) have been the subject of extensive study in the past \citep{1,2,3,4,5,6,7,8,9,10,11,12,13,14,15}. At finite temperatures, the BEC oscillates in the presence of non-condensate fraction and this leads to the damping of the low energy collective modes\citep{16,17,18,19,20,21,22,23}. The precise calculation of the damping is of much importance to understand experimental results and quantum many body physics of BEC.

Damping mechanism associated with collective excitations of Bose condensed atoms interacting with a non-condensed, thermal component is not well understood and still represents a challenging problem in theoretical physics. The damping mechanism of the collective modes depend on the temperature and density. There are two regimes, collisional and collisionless. The collisional regime is described by high temperature and densities and is understood in terms of the two-fluid hydrodynamics\citep{24,25,26,27,28,28a}. On the other hand low temperature and densities mark the collisionless regime. Damping in this regime is attributed to coupling between excitations and can be understood by mean-field approach\citep{29,30,31,32,33,34,35,36,37,39}.
Damping of the low-lying collective modes in the collisionless regime and finite temperature is of landau type (one quantum of oscillation being absorbed by a thermal excitation , which in turn produces another thermal excitation. In contrast at low temperatures Baliaev damping (process of decay of a quantum of condensate mode into two excitations with lower energy) becomes dominant.

Two-component BEC either as two overlapping atomic hyperfine states or as adjacent traps coupled by tunnel effects have also attracted much attention. Two-component Bose gases have been used to study phase coherence\citep{40}, Josephson physics\citep{41,42,43,44,45}, spin textures\citep{46,47,48,49}, quantum information processing\citep{50,51,52}, parametric excitation\citep{53} and polaritonic states\citep{54}, spin-charge separation\citep{55}.

In view of the interesting physics that emerges out of the coupling between two condensates, we attempt in this paper to understand the damping mechanism of such a system using mean-field approach in the collisionless regime based on an approach, where the equilibrium value of the anomalous density is neglected. We show that both for Landau and Baliaev damping, the excitations of the two-component are coupled (i.e collective mode of the coupled system is connected to thermal fluctuations of the condensate and non-condensate parts of both the components).

\section{Finite temperature Mean Field Model}

We consider a two-component Bose gas mixture at thermodynamic equilibrium at temperature $T$ in non-uniform external fields $V_{1}(\vec{r})$ and $V_{2}(\vec{r})$.The two components labeled by $\sigma={1,2}$ interact with each other and can exchange atoms to maintain chemical equilibrium. The total number of atoms, $N=N_{1}+N_{2}$, is conserved but the number of atoms in each component $N_{\sigma}$ is not. The system is described by the grand-canonical Hamiltonian which has the form:

\begin{eqnarray}
K &=& H-\mu N = \int d\vec{r} \ \psi_{1}^{\dagger}(\vec{r}) \left[ \frac{-\hbar \nabla^{2}}{2 m_{1}} + V_{1}(\vec{r})-\mu \right] \psi_{1}(\vec{r})+\int d\vec{r} \ \psi_{2}^{\dagger}(\vec{r}) \left[ \frac{-\hbar \nabla^{2}}{2 m_{2}} +V_{2}(\vec{r})-\mu \right] \psi_{2}(\vec{r})\nonumber \\
&+& \frac{g_{11}}{2} \int d\vec{r} \ \psi_{1}^{\dagger}(\vec{r})\psi_{1}^{\dagger}(\vec{r})\psi_{1}(\vec{r})\psi_{1}(\vec{r})
+\frac{g_{22}}{2} \int d\vec{r} \ \psi_{2}^{\dagger}(\vec{r})\psi_{2}^{\dagger}(\vec{r})\psi_{2}(\vec{r})\psi_{2}(\vec{r})
+g_{12} \int d\vec{r} \ \psi_{1}^{\dagger}(\vec{r})\psi_{1}(\vec{r})\psi_{2}^{\dagger}(\vec{r})\psi_{2}(\vec{r})
\end{eqnarray}

Here $\psi_{\sigma}^{\dagger}(\vec{r},t)$ and $\psi_{\sigma}(\vec{r},t)$ $(\sigma=1,2)$ are the creation and annihilation particle field operators for the two components. In the above equation $g_{\sigma,\sigma}$ is the intra-component interaction coupling strength and $g_{12}$ is the inter-component interaction coupling strength. Further, $m_{1}$ and $m_{2}$ are the masses of the atoms of the two-components. The chemical potential of the mixture is $\mu$.
The particle field operators for the two components satisfy the following two coupled equations of motion:

\begin{equation}
i\hbar \frac{\partial}{\partial t} \psi_{1}(\vec{r},t)= \left[ \frac{-\hbar \nabla^{2}}{2 m_{1}} + V_{1}(\vec{r})-\mu \right] \psi_{1}(\vec{r},t)
+g_{11}\psi_{1}^{\dagger}(\vec{r},t)\psi_{1}^{\dagger}(\vec{r},t)\psi_{1}(\vec{r},t)+g_{12}\psi_{2}^{\dagger}(\vec{r},t)\psi_{2}^{\dagger}(\vec{r},t)\psi_{1}(\vec{r},t)
\end{equation}

\begin{equation}
i\hbar \frac{\partial}{\partial t} \psi_{2}(\vec{r},t)= \left[ \frac{-\hbar \nabla^{2}}{2 m_{2}} + V_{2}(\vec{r})-\mu \right] \psi_{2}(\vec{r},t)
+g_{22}\psi_{2}^{\dagger}(\vec{r},t)\psi_{2}^{\dagger}(\vec{r},t)\psi_{2}(\vec{r},t)+g_{12}\psi_{1}^{\dagger}(\vec{r},t)\psi_{1}^{\dagger}(\vec{r},t)\psi_{2}(\vec{r},t)
\end{equation}

Non-equilibrium situations can be treated with the help of a time-dependent condensate wavefunction $\phi_{\sigma}(\vec{r},t)=<\psi_{\sigma}(\vec{r},t)>$, $(\sigma=1,2)$. Due to finite temperature, the particle field operators can be separated into a condensate and a non-condensate component,

\begin{equation}
\psi_{\sigma}(\vec{r},t)= \phi_{\sigma}(\vec{r},t)+\tilde{\psi}_{\sigma}(\vec{r},t).
\end{equation}

The non-condensate component satisfies the non-equilibrium average, $<\tilde{\psi}_{\sigma}(\vec{r},t)>=0$. We now apply the decomposition of Eqn.(4) to the equations of motion (2) and (3). We further assume that for a dilute Bose system, the non-equilibrium average of the cubic product of the non-condensate operators do not contribute to the dynamics of the system.

We then have the following equations of motion for $\phi_{1}(\vec{r},t)$ and $\phi_{2}(\vec{r},t)$:

\begin{eqnarray}
i\hbar \frac{\partial}{\partial t} \phi_{1}(\vec{r},t) &=& \left[ \frac{-\hbar \nabla^{2}}{2 m_{1}} + V_{1}(\vec{r})-\mu \right] \phi_{1}(\vec{r},t)
+g_{11} \left[ |\phi_{1}(\vec{r},t)|^{2} \phi_{1}(\vec{r},t) +2 \phi_{1}(\vec{r},t)\tilde{n}_{11}(\vec{r},t)+ \phi_{1}^{*}(\vec{r},t)\tilde{m}_{11}(\vec{r},t)   \right] \nonumber \\  &+& g_{12} \left[ |\phi_{2}(\vec{r},t)|^{2} \phi_{1}(\vec{r},t) +\phi_{2}^{*}(\vec{r},t)\tilde{m}_{21}(\vec{r},t)+ \phi_{2}(\vec{r},t)\tilde{n}_{21}(\vec{r},t)+\phi_{1}(\vec{r},t) \tilde{n}_{22}(\vec{r},t)   \right]
\end{eqnarray}

\begin{eqnarray}
i\hbar \frac{\partial}{\partial t} \phi_{2}(\vec{r},t) &=& \left[ \frac{-\hbar \nabla^{2}}{2 m_{2}} + V_{2}(\vec{r})-\mu \right] \phi_{2}(\vec{r},t)
+g_{22} \left[ |\phi_{2}(\vec{r},t)|^{2} \phi_{2}(\vec{r},t) +2 \phi_{2}(\vec{r},t)\tilde{n}_{22}(\vec{r},t)+ \phi_{2}^{*}(\vec{r},t)\tilde{m}_{22}(\vec{r},t)   \right] \nonumber \\  &+& g_{12} \left[ |\phi_{1}(\vec{r},t)|^{2} \phi_{2}(\vec{r},t) +\phi_{1}^{*}(\vec{r},t)\tilde{m}_{12}(\vec{r},t)+ \phi_{1}(\vec{r},t)\tilde{n}_{12}(\vec{r},t)+\phi_{2}(\vec{r},t) \tilde{n}_{11}(\vec{r},t)   \right]
\end{eqnarray}

In the above equations, we have introduced the following time dependent densities and cross-correlations:

\begin{eqnarray}
\tilde{n}_{\sigma,\sigma}(\vec{r},t)&=& <\tilde{\psi}_{\sigma}^{\dagger}(\vec{r},t) \tilde{\psi}_{\sigma}(\vec{r},t)> \nonumber \\
\tilde{m}_{\sigma,\sigma}(\vec{r},t)&=& <\tilde{\psi}_{\sigma}(\vec{r},t) \tilde{\psi}_{\sigma}(\vec{r},t)> \nonumber \\
\tilde{n}_{1,2}(\vec{r},t)&=& <\tilde{\psi}_{1}^{\dagger}(\vec{r},t) \tilde{\psi}_{2}(\vec{r},t)> \nonumber \\
\tilde{n}_{2,1}(\vec{r},t) &=& \tilde{n}_{1,2}^{\dagger}(\vec{r},t)= <\tilde{\psi}_{2}^{\dagger}(\vec{r},t) \tilde{\psi}_{1}(\vec{r},t)> \nonumber \\
\tilde{m}_{12}(\vec{r},t) &=& \tilde{m}_{21}(\vec{r},t)= <\tilde{\psi}_{1}(\vec{r},t) \tilde{\psi}_{2}(\vec{r},t)>
\end{eqnarray}

In order to study the small-amplitude dynamics of the system, the condensate is displaced from its stationary component $\phi_{\sigma}^{0}(\vec{r})$ by a small amount $\delta \phi_{\sigma}$. Consequently, the condensate part is decomposed into a stationary component and an excited component (small fluctuation) as

\begin{equation}
\phi_{\sigma}(\vec{r},t)= \phi_{\sigma}^{0}(\vec{r})+\delta \phi_{\sigma}(\vec{r},t)
\end{equation}

In a similar manner, we consider small fluctuations of the various densities and cross correlations of Eqn.(7) around their equilibrium values as:

\begin{eqnarray}
\tilde{n}_{\sigma,\sigma}(\vec{r},t)&=& \tilde{n}_{\sigma,\sigma}^{0}(\vec{r})+ \delta \bar{n}_{\sigma,\sigma}(\vec{r},t) \nonumber \\
\tilde{m}_{\sigma,\sigma}(\vec{r},t)&=&  \tilde{m}_{\sigma,\sigma}^{0}(\vec{r})+ \delta \bar{m}_{\sigma,\sigma}(\vec{r},t) \nonumber \\
\tilde{n}_{1,2}(\vec{r},t)&=&  \tilde{n}_{1,2}^{0}(\vec{r})+ \delta \bar{n}_{1,2}(\vec{r},t) \nonumber \\
\tilde{n}_{2,1}(\vec{r},t)&=&  \tilde{n}_{2,1}^{0}(\vec{r})+ \delta \bar{n}_{2,1}(\vec{r},t) \nonumber \\
\tilde{m}_{1,2}(\vec{r},t)&=&  \tilde{m}_{1,2}^{0}(\vec{r})+ \delta \bar{m}_{1,2}(\vec{r},t)
\end{eqnarray}

Here the equilibrium values are $\tilde{n}_{\sigma,\sigma}^{0}(\vec{r})=<\tilde{\psi}_{\sigma}^{\dagger}(\vec{r})\tilde{\psi}_{\sigma}(\vec{r})>_{0}$, $\tilde{m}_{\sigma,\sigma}^{0}(\vec{r})=<\tilde{\psi}_{\sigma}(\vec{r})\tilde{\psi}_{\sigma}(\vec{r})>_{0}$, $\tilde{n}_{1,2}^{0}(\vec{r})=<\tilde{\psi}_{1}^{\dagger}(\vec{r})\tilde{\psi}_{2}(\vec{r})>_{0}$, $\tilde{n}_{2,1}^{0}(\vec{r})=<\tilde{\psi}_{2}^{\dagger}(\vec{r})\tilde{\psi}_{1}(\vec{r})>_{0}$, $\tilde{m}_{1,2}^{0}(\vec{r})=<\tilde{\psi}_{1}(\vec{r})\tilde{\psi}_{2}(\vec{r})>_{0}$

We ignore the effects arising from the equilibrium values of the anomalous densities $\tilde{m}_{\sigma,\sigma}^{0}(\vec{r})$ and the cross correlations $\tilde{n}_{1,2}^{0}(\vec{r})$, $\tilde{n}_{2,1}^{0}(\vec{r})$ and $\tilde{m}_{1,2}^{0}(\vec{r})$.

Using these approximations, the time dependent equations for $\delta \phi_{1}(\vec{r},t)$ and $\delta \phi_{2}(\vec{r},t)$ is obtained after linearizing as

\begin{eqnarray}
i \hbar \frac{\partial}{\partial t}\delta \phi_{1}(\vec{r},t)&=& \left[\frac{-\hbar^2 \nabla^{2}}{2 m_{1}}+V_{1}(\vec{r})-\mu+2 g_{11}n_{1}(\vec{r})+g_{12}n_{2}(\vec{r}) \right] \delta \phi_{1}(\vec{r},t)+g_{11}n_{1}^{0}(\vec{r})\delta \phi_{1}^{*}(\vec{r},t)+2g_{11}\phi_{1}^{0}(\vec{r})\delta \bar{n}_{11}(\vec{r},t)\nonumber \\
&+& g_{11} \phi_{1}^{0}(\vec{r})\delta \bar{m}_{11}(\vec{r},t)+g_{12}\phi_{1}^{0}(\vec{r})\phi_{2}^{0}(\vec{r})\left[ \delta \phi_{2}(\vec{r},t)+\delta \phi_{2}^{*}(\vec{r},t) \right]+g_{12} \phi_{2}^{0} \left[ \delta \bar{m}_{21}+\delta \bar{n}_{21} \right]+g_{12} \phi_{1}^{0} \delta \bar{n}_{22},
\end{eqnarray}

\begin{eqnarray}
i \hbar \frac{\partial}{\partial t}\delta \phi_{2}(\vec{r},t)&=& \left[\frac{-\hbar^2 \nabla^{2}}{2 m_{2}}+V_{2}(\vec{r})-\mu+2 g_{22}n_{2}(\vec{r})+g_{12}n_{1}(\vec{r}) \right] \delta \phi_{2}(\vec{r},t)+g_{22}n_{2}^{0}(\vec{r})\delta \phi_{2}^{*}(\vec{r},t)+2g_{22}\phi_{2}^{0}(\vec{r})\delta \bar{n}_{22}(\vec{r},t)\nonumber \\
&+& g_{22} \phi_{2}^{0}(\vec{r})\delta \bar{m}_{22}(\vec{r},t)+g_{12}\phi_{1}^{0}(\vec{r})\phi_{2}^{0}(\vec{r})\left[ \delta \phi_{1}(\vec{r},t)+\delta \phi_{1}^{*}(\vec{r},t) \right]+g_{12} \phi_{1}^{0} \left[ \delta \bar{m}_{12}+\delta \bar{n}_{12} \right]+g_{12} \phi_{2}^{0} \delta \bar{n}_{11}.
\end{eqnarray}

Note from Eqns.(10) and (11), the fluctuations of each component is coupled to the condensate and non-condensate fluctuations of the other component. In order to obtain equations of motion for $\delta \bar{n}_{\sigma,\sigma}(\vec{r})$ and $\delta \bar{m}_{\sigma,\sigma}(\vec{r})$, we first introduce the Bogoliubov transformations

\begin{eqnarray}
\tilde{\psi}_{\sigma}(\vec{r},t)&=& \sum_{j} \left[u_{\sigma,j}(\vec{r})\alpha_{j}(t)+v_{\sigma,j}^{*}(\vec{r})\alpha_{j}^{\dagger}(t)\right], \nonumber \\
\tilde{\psi}_{\sigma}^{\dagger}(\vec{r},t)&=& \sum_{j} \left[u_{\sigma,j}^{*}(\vec{r})\alpha_{j}^{\dagger}(t)+v_{\sigma,j}(\vec{r})\alpha_{j}(t)\right],
\end{eqnarray}

with normalization of the Bogoliubov amplitudes $u_{\sigma,j}$ and $v_{\sigma,j}$,

\begin{equation}
\int d\vec{r} \left[u_{\sigma,j}^{*}(\vec{r})u_{\sigma,j}(\vec{r})-v_{\sigma,j}^{*}(\vec{r})v_{\sigma,j}(\vec{r})\right]=\delta_{ij},
\end{equation}

and $[\alpha_{i}^{\dagger}\alpha_{i}]=\delta_{ij}$. Here $\alpha_{j}$ and $\alpha_{j}^{\dagger}$ are quasi-particle operators. We further introduce the following quasi-particle distribution functions which would help us in calculating
$\delta \bar{n}_{\sigma,\sigma}(\vec{r},t)$ and $\delta \bar{m}_{\sigma,\sigma}(\vec{r},t)$,

\begin{eqnarray}
f_{ij}(t)&=& <\alpha_{i}^{\dagger}(t) \alpha_{j}(t)>, \nonumber \\
g_{ij}(t)&=& <\alpha_{i}(t) \alpha_{j}(t)>.
\end{eqnarray}

The distribution functions obey the following equations of motion

\begin{eqnarray}
i \hbar \frac{\partial}{\delta \partial} f_{ij}(t)= <[\alpha_{i}^{\dagger}(t) \alpha_{j}(t), K]>, \nonumber \\
i \hbar \frac{\partial}{\delta \partial} g_{ij}(t)= <[\alpha_{i}(t) \alpha_{j}(t), K]>.
\end{eqnarray}

We now need to write the grand-canonical Hamiltonian with the relevant terms. To this end, we note that only terms quadratic and quartic in the non-condensate operators $\tilde{\psi}_{\sigma}$, $\tilde{\psi}_{\sigma}^{\dagger}$ give nonzero contributions. We also keep the Hamiltonian linear in the fluctuations $\delta \phi_{\sigma}(\vec{r},t)$.

\begin{equation}
K=K_{2}^{(0)}+K_{2}^{(1)}+K_{4}^{(0)},
\end{equation}

where,

\begin{eqnarray}
K_{2}^{(0)}&=& \int d \vec{r} \left[\tilde{\psi}_{1}^{\dagger}(\vec{r},t)\left( \frac{-\hbar^{2} \nabla^{2}}{2 m_{1}}+V_{1}(\vec{r})-\mu +2g_{11}n_{1}^{0}(\vec{r})+g_{12}n_{2}^{0}(\vec{r}) \right) \tilde{\psi}_{1}(\vec{r},t)\nonumber
+\frac{g_{11}}{2}n_{1}^{0}(\vec{r})\left(\tilde{\psi}_{1}^{\dagger}(\vec{r},t)\tilde{\psi}_{1}^{\dagger}(\vec{r},t)+\tilde{\psi}_{1}(\vec{r},t)\tilde{\psi}_{1}(\vec{r},t)\right)                       \right]\nonumber \\
&+& \int d \vec{r} \left[\tilde{\psi}_{2}^{\dagger}(\vec{r},t)\left( \frac{-\hbar^{2} \nabla^{2}}{2 m_{2}}+V_{2}(\vec{r})-\mu +2g_{22}n_{2}^{0}(\vec{r})+g_{12}n_{1}^{0}(\vec{r}) \right) \tilde{\psi}_{2}(\vec{r},t)\nonumber
+\frac{g_{22}}{2}n_{2}^{0}(\vec{r})\left(\tilde{\psi}_{2}^{\dagger}(\vec{r},t)\tilde{\psi}_{2}^{\dagger}(\vec{r},t)+\tilde{\psi}_{2}(\vec{r},t)\tilde{\psi}_{2}(\vec{r},t)\right)                       \right] \nonumber \\
&+& g_{12} \int d \vec{r} \ \phi_{1}^{0}(\vec{r}) \phi_{2}^{0}(\vec{r})\left[\tilde{\psi}_{1}(\vec{r},t)\tilde{\psi}_{2}(\vec{r},t) +\tilde{\psi}_{1}^{\dagger}(\vec{r},t)\tilde{\psi}_{2}^{\dagger}(\vec{r},t)+ \tilde{\psi}_{1}(\vec{r},t) \tilde{\psi}_{2}^{\dagger}(\vec{r},t)+\tilde{\psi}_{1}^{\dagger}(\vec{r},t)\tilde{\psi}_{2}(\vec{r},t)\right],
\end{eqnarray}

\begin{eqnarray}
K_{2}^{(1)}&=& \sum_{\sigma=1,2}\int d \vec{r} \ [ 2 g_{\sigma,\sigma}\phi_{\sigma}^{0}(\vec{r}) \left( \delta \phi_{\sigma}(\vec{r},t)+\delta \phi_{\sigma}^{*}(\vec{r},t)  \right)\tilde{\psi}_{\sigma}^{\dagger}(\vec{r},t)\tilde{\psi}_{\sigma}(\vec{r},t)\nonumber \\
  &+& g_{\sigma,\sigma} \phi_{\sigma}^{0}(\vec{r}) \left( \delta \phi_{\sigma}(\vec{r},t)\tilde{\psi}_{\sigma}^{\dagger}(\vec{r},t)\tilde{\psi}_{\sigma}^{\dagger}(\vec{r},t)+ \delta \phi_{\sigma}(\vec{r},t)^{*} \tilde{\psi}_{\sigma}(\vec{r},t)\tilde{\psi}_{\sigma}(\vec{r},t)\right)] \nonumber \\
  &+& g_{12} \int d \vec{r} \ [\phi_{1}^{0}(\vec{r}) \left( \delta \phi_{1}(\vec{r},t)+\delta \phi_{1}^{*}(\vec{r},t)\right)\tilde{\psi}_{2}^{\dagger}(\vec{r},t)\tilde{\psi}_{2}(\vec{r},t)+\phi_{2}^{0}(\vec{r}) \left( \delta \phi_{2}(\vec{r},t)+\delta \phi_{2}^{*}(\vec{r},t)\right)\tilde{\psi}_{1}^{\dagger}(\vec{r},t)\tilde{\psi}_{1}(\vec{r},t)] \nonumber \\
  &+& g_{12} \int d \vec{r} \ [\left( \phi_{1}^{0}(\vec{r})  \delta \phi_{2}^{*}(\vec{r},t)+\phi_{2}^{0}(\vec{r}) \delta \phi_{1}^{*}(\vec{r},t)\right)\tilde{\psi}_{1}(\vec{r},t)\tilde{\psi}_{2}(\vec{r},t)+ \left( \phi_{1}^{0}(\vec{r})  \delta \phi_{2}(\vec{r},t)+\phi_{2}^{0}(\vec{r}) \delta \phi_{1}(\vec{r},t)\right)\tilde{\psi}_{1}^{\dagger}(\vec{r},t)\tilde{\psi}_{2}^{\dagger}(\vec{r},t) ]\nonumber \\
  &+& g_{12} \int d \vec{r} \ [\left( \phi_{1}^{0}(\vec{r})  \delta \phi_{2}(\vec{r},t)+\phi_{2}^{0}(\vec{r}) \delta \phi_{1}^{*}(\vec{r},t)\right)\tilde{\psi}_{1}(\vec{r},t)\tilde{\psi}_{2}^{\dagger}(\vec{r},t)+ \left( \phi_{1}^{0}(\vec{r})  \delta \phi_{2}^{*}(\vec{r},t)+\phi_{2}^{0}(\vec{r}) \delta \phi_{1}(\vec{r},t)\right)\tilde{\psi}_{1}^{\dagger}(\vec{r},t)\tilde{\psi}_{2}(\vec{r},t) ], \nonumber
  \\
\end{eqnarray}

\begin{eqnarray}
K_{4}^{0}=2 g_{11} \int d \vec{r}\ \tilde{n}_{11}^{0}(\vec{r}) \tilde{\psi}_{1}^{\dagger}(\vec{r},t)\tilde{\psi}_{1}(\vec{r},t)+2 g_{22} \int d \vec{r}\ \tilde{n}_{22}^{0} (\vec{r}) \tilde{\psi}_{2}^{\dagger}(\vec{r},t)\tilde{\psi}_{2}(\vec{r},t).
\end{eqnarray}

In the interaction terms, we have neglected the linear in the fluctuations $\delta \bar{n}_{\sigma,\sigma}$ and $\delta \bar{m}_{\sigma,\sigma}$. If the density of the non-condensate components is much less than the densities of the condensate components, the coupling to the condensate is more significant than the coupling to $\delta \bar{n}_{\sigma,\sigma}$ and $\delta \bar{m}_{\sigma,\sigma}$. In the following, we will only consider the case, $m_{1}$ $=$ $m_{2}$ $=$ $m$ and $V_{1}$ $=$ $V_{2}$ $=$ $V$.The operator $K_{2}^{0}+K_{4}^{0}$ can be diagonalized, in $\alpha_{j}$, $\alpha_{j}^{\dagger}$ if the functions $u_{\sigma,j}$ and $v_{\sigma,j}$, satisfy the following coupled BdG equations,

\begin{equation}
L_{1}u_{1j}(\vec{r})+g_{11}n_{1}^{0}(\vec{r})v_{1j}(\vec{r})+g_{12}\sqrt{n_{1}^{0}(\vec{r})}\sqrt{n_{2}^{0}(\vec{r})}(u_{2j}(\vec{r})+v_{2j}(\vec{r}))=\epsilon_{j} u_{1j}(\vec{r}),
\end{equation}

\begin{equation}
L_{1}v_{1j}(\vec{r})+g_{11}n_{1}^{0}(\vec{r})u_{1j}(\vec{r})+g_{12}\sqrt{n_{1}^{0}(\vec{r})}\sqrt{n_{2}^{0}(\vec{r})}(u_{2j}(\vec{r})+v_{2j}(\vec{r}))=-\epsilon_{j} v_{1j}(\vec{r}),
\end{equation}

\begin{equation}
L_{2}u_{2j}(\vec{r})+g_{22}n_{2}^{0}(\vec{r})v_{2j}(\vec{r})+g_{12}\sqrt{n_{1}^{0}(\vec{r})}\sqrt{n_{2}^{0}(\vec{r})}(u_{1j}(\vec{r})+v_{1j}(\vec{r}))=\epsilon_{j} u_{2j}(\vec{r}),
\end{equation}

\begin{equation}
L_{2}v_{2j}(\vec{r})+g_{22}n_{2}^{0}(\vec{r})u_{2j}(\vec{r})+g_{12}\sqrt{n_{1}^{0}(\vec{r})}\sqrt{n_{2}^{0}(\vec{r})}(u_{1j}(\vec{r})+v_{1j}(\vec{r}))=-\epsilon_{j} v_{2j}(\vec{r}),
\end{equation}

\begin{equation}
L_{1}=\frac{-\hbar^2 \nabla^{2}}{2m}+V(\vec{r})-\mu+2g_{11}n_{1}^{0}+g_{12}n_{2}^{0},
\end{equation}

\begin{equation}
L_{2}=\frac{-\hbar^2 \nabla^{2}}{2m}+V(\vec{r})-\mu+2g_{22}n_{2}^{0}+g_{12}n_{1}^{0}.
\end{equation}

The relevant Hamiltonian is,

\begin{equation}
K_{eff}=\sum_{j} \epsilon_{j}(\vec{r}) \alpha_{j}^{\dagger}(\vec{r}) \alpha_{j}+K_{2}^{(1)},
\end{equation}

where the quasi-particle energies $\epsilon_{j}$ are obtained by solving the BdG equations. Now to lowest order in the fluctuations, the equations of motion of the quasi-particle distribution $f_{ij}(t)$ and $g_{ij}(t)$ are,

\begin{eqnarray}
i \hbar \frac{\partial}{\partial t}f_{ij}(t)&=&(\epsilon_{j}-\epsilon_{i})f_{ij}(t)+(f_{i}^{0}-f_{j}^{0})[\int d\vec{r}\ [2g_{11}\phi_{1}^{0}(\vec{r})(\delta\phi_{1}(\vec{r},t)+\delta \phi_{1}^{*}(\vec{r},t))\nonumber \\
&+& g_{12}\phi_{2}^{0}(\vec{r})(\delta\phi_{2}(\vec{r},t)+\delta \phi_{2}^{*}(\vec{r},t))](u_{1i}(\vec{r})u_{1j}^{*}(\vec{r})+v_{1i}(\vec{r})v_{1j}^{*}(\vec{r}))\nonumber \\
&+& 2 g_{11} \int d \vec{r} \phi_{1}^{0}[\delta \phi_{1}(\vec{r},t)v_{1i}(\vec{r})u_{1j}^{*}(\vec{r})+\delta \phi_{1}^{*}(\vec{r},t)u_{1i}(\vec{r})v_{1j}^{*}(\vec{r})]\nonumber \\
&+& \int d\vec{r}\ [2g_{22}\phi_{2}^{0}(\vec{r})(\delta\phi_{2}(\vec{r},t)+\delta \phi_{2}^{*}(\vec{r},t))\nonumber \\
&+& g_{12}\phi_{1}^{0}(\vec{r})(\delta\phi_{1}(\vec{r},t)+\delta \phi_{1}^{*}(\vec{r},t))](u_{2i}(\vec{r})u_{2j}^{*}(\vec{r})+v_{2i}(\vec{r})v_{2j}^{*}(\vec{r}))\nonumber \\
&+& 2 g_{22} \int d \vec{r}\ \phi_{2}^{0}[\delta \phi_{2}(\vec{r},t)v_{2i}(\vec{r})u_{2j}^{*}(\vec{r})+\delta \phi_{2}^{*}(\vec{r},t)u_{2i}(\vec{r})v_{2j}^{*}(\vec{r})] \nonumber \\
&+& g_{12} \int d \vec{r} \ [( \phi_{1}^{0}(\vec{r})  \delta \phi_{2}^{*}(\vec{r},t)+\phi_{2}^{0}(\vec{r}) \delta \phi_{1}^{*}(\vec{r},t)) (u_{1i}(\vec{r})v_{2j}^{*}(\vec{r})+v_{1j}^{*}(\vec{r})u_{2i}(\vec{r}))\nonumber \\
 &+&\left( \phi_{1}^{0}(\vec{r})  \delta \phi_{2}(\vec{r},t)+\phi_{2}^{0}(\vec{r}) \delta \phi_{1}(\vec{r},t)\right)(u_{1j}^{*}(\vec{r})v_{2i}(\vec{r})+v_{1i}(\vec{r})u_{2j}^{*}(\vec{r}))] \nonumber \\
 &+& g_{12} \int d \vec{r} \ [( \phi_{1}^{0}(\vec{r})  \delta \phi_{2}(\vec{r},t)+\phi_{2}^{0}(\vec{r}) \delta \phi_{1}^{*}(\vec{r},t)) (u_{1i}(\vec{r})u_{2j}^{*}(\vec{r})+v_{1j}^{*}(\vec{r})v_{2i}(\vec{r}))\nonumber \\
 &+&\left( \phi_{1}^{0}(\vec{r})  \delta \phi_{2}^{*}(\vec{r},t)+\phi_{2}^{0}(\vec{r}) \delta \phi_{1}(\vec{r},t)\right)(u_{1j}^{*}(\vec{r})u_{2i}(\vec{r})+v_{1i}(\vec{r})v_{2j}^{*}(\vec{r}))] ]\nonumber \\
\end{eqnarray}

\begin{eqnarray}
i \hbar \frac{\partial}{\partial t}g_{ij}(t)&=&(\epsilon_{j}+\epsilon_{i})f_{ij}(t)+(1+f_{i}^{0}+f_{j}^{0})[\int d\vec{r}\ [2g_{11}\phi_{1}^{0}(\vec{r})(\delta\phi_{1}(\vec{r},t)+\delta \phi_{1}^{*}(\vec{r},t))\nonumber \\
&+& g_{12}\phi_{2}^{0}(\vec{r})(\delta\phi_{2}(\vec{r},t)+\delta \phi_{2}^{*}(\vec{r},t))](u_{1i}^{*}(\vec{r})v_{1j}^{*}(\vec{r})+v_{1i}^{*}(\vec{r})v_{1j}^{*}(\vec{r}))\nonumber \\
&+& 2 g_{11} \int d \vec{r} \phi_{1}^{0}[\delta \phi_{1}(\vec{r},t)u_{1i}^{*}(\vec{r})u_{1j}^{*}(\vec{r})+\delta \phi_{1}^{*}(\vec{r},t)v_{1i}^{*}(\vec{r})v_{1j}^{*}(\vec{r})]\nonumber \\
&+& \int d\vec{r}\ [2g_{22}\phi_{2}^{0}(\vec{r})(\delta\phi_{2}(\vec{r},t)+\delta \phi_{2}^{*}(\vec{r},t))\nonumber \\
&+& g_{12}\phi_{1}^{0}(\vec{r})(\delta\phi_{1}(\vec{r},t)+\delta \phi_{1}^{*}(\vec{r},t))](u_{2i}^{*}(\vec{r})v_{2j}^{*}(\vec{r})+v_{2i}^{*}(\vec{r})u_{2j}^{*}(\vec{r}))\nonumber \\
&+& 2 g_{22} \int d \vec{r} \ \phi_{2}^{0}[\delta \phi_{2}(\vec{r},t)u_{2i}^{*}(\vec{r})u_{2j}^{*}(\vec{r})+\delta \phi_{2}^{*}(\vec{r},t)v_{2i}^{*}(\vec{r})v_{2j}^{*}(\vec{r})] \nonumber \\
&+& g_{12} \int d \vec{r} \ [( \phi_{1}^{0}(\vec{r})  \delta \phi_{2}^{*}(\vec{r},t)+\phi_{2}^{0}(\vec{r}) \delta \phi_{1}^{*}(\vec{r},t)) (v_{1i}^{*}(\vec{r})v_{2j}^{*}(\vec{r}))\nonumber \\
 &+&\left( \phi_{1}^{0}(\vec{r})  \delta \phi_{2}(\vec{r},t)+\phi_{2}^{0}(\vec{r}) \delta \phi_{1}(\vec{r},t)\right)(u_{1i}^{*}(\vec{r})u_{2j}^{*}(\vec{r}))] \nonumber \\
 &+& g_{12} \int d \vec{r} \ [( \phi_{1}^{0}(\vec{r})  \delta \phi_{2}(\vec{r},t)+\phi_{2}^{0}(\vec{r}) \delta \phi_{1}^{*}(\vec{r},t)) (v_{1i}^{*}(\vec{r})u_{2j}^{*}(\vec{r}))\nonumber \\
 &+&\left( \phi_{1}^{0}(\vec{r})  \delta \phi_{2}^{*}(\vec{r},t)+\phi_{2}^{0}(\vec{r}) \delta \phi_{1}(\vec{r},t)\right)(u_{1i}^{*}(\vec{r})v_{2j}^{*}(\vec{r}))] ]\nonumber \\
\end{eqnarray}

In the above equations, $f_{ij}$ is the equilibrium density of the quasi-particles. The normal and anomalous quasi-particle densities $\delta \bar{n}_{\sigma \sigma}$, $\delta \bar{m}_{\sigma \sigma}$ and cross correlations $\delta \bar{n}_{21}$, $\delta \bar{n}_{12}$, $\delta \bar{m}_{21}$, $\delta \bar{m}_{12}$, can be written in terms of $f_{ij}(t)$ and $g_{ij}(t)$.

\begin{eqnarray}
i \hbar \frac{\partial}{\partial t}\delta \phi_{1}(\vec{r},t)&=& \left[\frac{-\hbar^2 \nabla^{2}}{2 m_{1}}+V_{1}(\vec{r})-\mu+2 g_{11}n_{1}(\vec{r})+g_{12}n_{2}(\vec{r}) \right] \delta \phi_{1}(\vec{r},t)+g_{11}n_{1}^{0}(\vec{r})\delta \phi_{1}^{*}(\vec{r},t)\nonumber \\
&+& g_{12} \sqrt{n_{1}^{0}(\vec{r})}\sqrt{n_{2}^{0}(\vec{r})}(\delta \phi_{2}^{*}(\vec{r},t)+\delta \phi_{2}(\vec{r},t))\nonumber \\
&+& g_{11} \phi_{1}^{0} \sum_{ij} [ 2[u_{1i}^{*}(\vec{r})u_{1j}(\vec{r})+v_{1i}^{*}(\vec{r})v_{1j}(\vec{r})+v_{1i}^{*}(\vec{r})u_{1j}(\vec{r}) ]f_{ij}(t)\nonumber \\
&+& [2v_{1i}(\vec{r})u_{1j}(\vec{r})+u_{1i}(\vec{r})u_{1j}(\vec{r})]g_{ij}(t)+[2u_{1i}^{*}(\vec{r})v_{1j}^{*}(\vec{r})+v_{1i}^{*}(\vec{r})v_{1j}^{*}(\vec{r})]g_{ij}^{*}(t)]\nonumber \\
&+& g_{12} \phi_{2}^{0}[[u_{2j}(\vec{r})v_{1i}^{*}(\vec{r})+v_{2i}^{*}(\vec{r})u_{1j}(\vec{r})+u_{2i}^{*}(\vec{r})u_{1j}(\vec{r})+v_{2j}(\vec{r})v_{1i}^{*}(\vec{r})]f_{ij}(t)\nonumber \\
&+&[v_{2i}(\vec{r})u_{1j}(\vec{r})+u_{2i}(\vec{r})u_{1j}]g_{ij}(t)+[u_{2i}^{*}(\vec{r})v_{1j}^{*}(\vec{r})+v_{2i}^{*}(\vec{r})v_{1j}^{*}]g_{ij}^{*}(t)] \nonumber \\
&+& g_{12} \phi_{1}^{0}[(u_{2i}^{*}(\vec{r})u_{2j}(\vec{r})+v_{2j}(\vec{r})v_{2i}^{*}(\vec{r}))f_{ij}(t)+u_{2i}^{*}(t)v_{2j}^{*}(\vec{r})g_{ij}^{*}(t)+v_{2i}(\vec{r})u_{2j}(\vec{r})g_{ij}(t)]
\end{eqnarray}

\begin{eqnarray}
i \hbar \frac{\partial}{\partial t}\delta \phi_{2}(\vec{r},t)&=& \left[\frac{-\hbar^2 \nabla^{2}}{2 m_{2}}+V_{2}(\vec{r})-\mu+2 g_{22}n_{2}(\vec{r})+g_{12}n_{1}(\vec{r}) \right] \delta \phi_{2}(\vec{r},t)+g_{22}n_{2}^{0}(\vec{r})\delta \phi_{2}^{*}(\vec{r},t)\nonumber \\
&+& g_{12} \sqrt{n_{1}^{0}(\vec{r})}\sqrt{n_{2}^{0}(\vec{r})}(\delta \phi_{1}^{*}(\vec{r},t)+\delta \phi_{1}(\vec{r},t))\nonumber \\
&+& g_{22} \phi_{2}^{0} \sum_{ij} [ 2[u_{2i}^{*}(\vec{r})u_{2j}(\vec{r})+v_{2i}^{*}(\vec{r})v_{2j}(\vec{r})+v_{2i}^{*}(\vec{r})u_{2j}(\vec{r}) ]f_{ij}(t)\nonumber \\
&+& [2v_{2i}(\vec{r})u_{2j}(\vec{r})+u_{2i}(\vec{r})u_{2j}(\vec{r})]g_{ij}(t)+[2u_{2i}^{*}(\vec{r})v_{2j}^{*}(\vec{r})+v_{2i}^{*}(\vec{r})v_{2j}^{*}(\vec{r})]g_{ij}^{*}(t)]\nonumber \\
&+& g_{12} \phi_{1}^{0}[[u_{1j}(\vec{r})v_{2i}^{*}(\vec{r})+v_{1i}^{*}(\vec{r})u_{2j}(\vec{r})+u_{1i}^{*}(\vec{r})u_{2j}(\vec{r})+v_{1j}(\vec{r})v_{2i}^{*}(\vec{r})]f_{ij}(t)\nonumber \\
&+&[u_{1i}(\vec{r})u_{2j}(\vec{r})+v_{1i}(\vec{r})u_{2j}]g_{ij}(t)+[v_{1i}^{*}(\vec{r})v_{2j}^{*}(\vec{r})+u_{1i}^{*}(\vec{r})v_{2j}^{*}]g_{ij}^{*}(t)] \nonumber \\
&+& g_{12} \phi_{2}^{0}[(u_{1i}^{*}(\vec{r})u_{1j}(\vec{r})+v_{1j}(\vec{r})v_{1i}^{*}(\vec{r}))f_{ij}(t)+u_{1i}^{*}(t)v_{1j}^{*}(\vec{r})g_{ij}^{*}(t)+v_{1i}(\vec{r})u_{1j}(\vec{r})g_{ij}(t)]
\end{eqnarray}

Eqns.(27)-(30) represent the small amplitude oscillations of the double condensate coupled to the non-condensate particles in the collisionless regime.

Let us suppose that the two components of the condensate oscillates with frequency $\omega$:

$\delta \phi_{1}(\vec{r},t)=\delta \xi_{1}(\vec{r}) e^{-i\omega t}$,$\delta \phi_{1}^{*}(\vec{r},t)=\delta \xi_{2}(\vec{r}) e^{-i\omega t}$,$\delta \phi_{2}(\vec{r},t)=\delta \xi_{3}(\vec{r}) e^{-i\omega t}$,$\delta \phi_{2}^{*}(\vec{r},t)=\delta \xi_{4}(\vec{r}) e^{-i\omega t}$.

The Fourier transform of $f_{ij}(t)$ and $g_{ij}(t)$ at the driving frequency $\omega$ is given by

\begin{eqnarray}
f_{ij}(\omega)&=&\frac{f_{i}^{0}-f_{j}^{0}}{\hbar \omega+(\epsilon_{i}-\epsilon_{j})+i0}\int d \vec{r} \ [2g_{11}\phi_{1}^{0}[\delta \xi_{1}(u_{1i}u_{1j}^{*}+v_{1i}v_{1j}^{*}+v_{1i}u_{1j}^{*})+\delta \xi_{2}(u_{1i}u_{1j}^{*}+v_{1i}v_{1j}^{*}+u_{1i}v_{1j}^{*})]\nonumber \\
&+& 2g_{22}\phi_{2}^{0}[\delta \xi_{3}(u_{2i}u_{2j}^{*}+v_{2i}v_{2j}^{*}+v_{2i}u_{2j}^{*})+\delta \xi_{4}(u_{2i}u_{2j}^{*}+v_{2i}v_{2j}^{*}+u_{2i}v_{2j}^{*})] \nonumber \\
&+& g_{12} \phi_{1}^{0}(\delta \xi_{1}+\delta \xi_{2})(u_{2i}u_{2j}^{*}+v_{2i}v_{2j}^{*})+g_{12} \phi_{2}^{0}(\delta \xi_{3}+\delta \xi_{4})(u_{1i}u_{1j}^{*}+v_{1i}v_{1j}^{*})\nonumber \\
&+& g_{12}[(\phi_{1}^{0}\delta \xi_{4}+\phi_{2}^{0} \delta \xi_{2})(v_{1j}^{*}u_{2i}+u_{1i}v_{2j}^{*})+(\phi_{1}^{0}\delta \xi_{3}+\phi_{2}^{0} \delta \xi_{1})(v_{1i}u_{2j}^{*}+v_{2i}u_{1j}^{*})]\nonumber \\
&+& g_{12}[(\phi_{1}^{0}\delta \xi_{3}+\phi_{2}^{0} \delta \xi_{2})(u_{1i}u_{2j}^{*}+v_{1j}^{*}v_{2i})+(\phi_{1}^{0}\delta \xi_{4}+\phi_{2}^{0} \delta \xi_{1})(v_{1i}v_{2j}^{*}+u_{2i}u_{1j}^{*})]]
\end{eqnarray}

\begin{eqnarray}
g_{ij}(\omega)&=&\frac{1+f_{i}^{0}+f_{j}^{0}}{\hbar \omega-(\epsilon_{i}+\epsilon_{j})+i0}\int d \vec{r} \ [2g_{11}\phi_{1}^{0}[\delta \xi_{1}(u_{1i}^{*}v_{1j}^{*}+v_{1i}^{*}u_{1j}^{*}+u_{1i}^{*}u_{1j}^{*})+\delta \xi_{2}(u_{1i}^{*}v_{1j}^{*}+v_{1i}^{*}u_{1j}^{*}+v_{1i}^{*}v_{1j}^{*})]\nonumber \\
&+& 2g_{22}\phi_{2}^{0}[\delta \xi_{3}(u_{2i}^{*}v_{2j}^{*}+v_{2i}^{*}u_{2j}^{*}+u_{2i}^{*}u_{2j}^{*})+\delta \xi_{4}(u_{2i}^{*}v_{2j}^{*}+v_{2i}^{*}u_{2j}^{*}+v_{2i}^{*}v_{2j}^{*})]\nonumber \\
&+& g_{12} \phi_{1}^{0}(\delta \xi_{1}+\delta \xi_{2})(u_{2i}^{*}v_{2j}^{*}+v_{2i}^{*}u_{2j}^{*})+g_{12} \phi_{2}^{0}(\delta \xi_{3}+\delta \xi_{4})(u_{1i}^{*}v_{1j}^{*}+v_{1i}^{*}u_{1j}^{*})\nonumber \\
&+& g_{12}[(\phi_{1}^{0}\delta \xi_{4}+\phi_{2}^{0} \delta \xi_{2})(v_{1i}^{*}v_{2j}^{*})+(\phi_{1}^{0}\delta \xi_{3}+\phi_{2}^{0} \delta \xi_{1})(u_{1i}^{*}u_{2j}^{*})]\nonumber \\
&+& g_{12}[(\phi_{1}^{0}\delta \xi_{3}+\phi_{2}^{0} \delta \xi_{2})(v_{1i}^{*}u_{2j}^{*})+(\phi_{1}^{0}\delta \xi_{4}+\phi_{2}^{0} \delta \xi_{1})(u_{1i}^{*}v_{2j}^{*})]]
\end{eqnarray}

We can get the energy correction to the collective mode of the coupled system in the Landau and Baliaev mechanisms by using the Fourier transformation of Eqns. (31) and (32),

\begin{equation}
\hbar \omega=\hbar \omega_{0}+\sum_{ij}(f_{ij}^{0}-f_{j}^{0}) \frac{|A_{ij}|^{2}}{\hbar \omega_{0}+(\epsilon_{i}-\epsilon_{j})+i0}+\sum_{ij}(1+f_{ij}^{0}+f_{j}^{0}) \frac{|B_{ij}|^{2}}{\hbar \omega_{0}-(\epsilon_{i}+\epsilon_{j})+i0},
\end{equation}

where $\omega_{0}$ is the unperturbed eigenfrequency and $\delta \xi_{l}^{0}(l=1,2,3,4)$ are the unperturbed condensate amplitudes of the two component Bose gas.

\begin{eqnarray}
A_{ij}&=&\int d\vec{r}\ [2g_{11}\phi_{1}^{0}[\delta \xi_{1}^{0}(u_{1i}u_{1j}^{*}+v_{1i}v_{1j}^{*}+v_{1i}u_{1j}^{*})+\delta \xi_{2}^{0}(u_{1i}u_{1j}^{*}+v_{1i}v_{1j}^{*}+u_{1i}v_{1j}^{*})]\nonumber \\
&+& 2g_{22}\phi_{2}^{0}[\delta \xi_{3}^{0}(u_{2i}u_{2j}^{*}+v_{2i}v_{2j}^{*}+v_{2i}u_{2j}^{*})+\delta \xi_{4}^{0}(u_{2i}u_{2j}^{*}+v_{2i}v_{2j}^{*}+u_{2i}v_{2j}^{*})] \nonumber \\
&+& g_{12} \phi_{1}^{0}(\delta \xi_{1}^{0}+\delta \xi_{2}^{0})(u_{2i}u_{2j}^{*}+v_{2i}v_{2j}^{*})+g_{12} \phi_{2}^{0}(\delta \xi_{3}^{0}+\delta \xi_{4}^{0})(u_{1i}u_{1j}^{*}+v_{1i}v_{1j}^{*})\nonumber \\
&+& g_{12}[(\phi_{1}^{0}\delta \xi_{4}^{0}+\phi_{2}^{0} \delta \xi_{2}^{0})(v_{1j}^{*}u_{2i}+u_{1i}v_{2j}^{*})+(\phi_{1}^{0}\delta \xi_{3}^{0}+\phi_{2}^{0} \delta \xi_{1}^{0})(v_{1i}u_{2j}^{*}+v_{2i}u_{1j}^{*})]\nonumber \\
&+& g_{12}[(\phi_{1}^{0}\delta \xi_{3}^{0}+\phi_{2}^{0} \delta \xi_{2}^{0})(u_{1i}u_{2j}^{*}+v_{1j}^{*}v_{2i})+(\phi_{1}^{0}\delta \xi_{4}^{0}+\phi_{2}^{0} \delta \xi_{1}^{0})(v_{1i}v_{2j}^{*}+u_{2i}u_{1j}^{*})]]
\end{eqnarray}

\begin{eqnarray}
B_{ij}&=&\int d \vec{r} \ [2g_{11}\phi_{1}^{0}[\delta \xi_{1}^{0}(u_{1i}^{*}v_{1j}^{*}+v_{1i}^{*}u_{1j}^{*}+u_{1i}^{*}u_{1j}^{*})+\delta \xi_{2}^{0}(u_{1i}^{*}v_{1j}^{*}+v_{1i}^{*}u_{1j}^{*}+v_{1i}^{*}v_{1j}^{*})]\nonumber \\
&+& 2g_{22}\phi_{2}^{0}[\delta \xi_{3}^{0}(u_{2i}^{*}v_{2j}^{*}+v_{2i}^{*}u_{2j}^{*}+u_{2i}^{*}u_{2j}^{*})+\delta \xi_{4}^{0}(u_{2i}^{*}v_{2j}^{*}+v_{2i}^{*}u_{2j}^{*}+v_{2i}^{*}v_{2j}^{*})]\nonumber \\
&+& g_{12} \phi_{1}^{0}(\delta \xi_{1}^{0}+\delta \xi_{2}^{0})(u_{2i}^{*}v_{2j}^{*}+v_{2i}^{*}u_{2j}^{*})+g_{12} \phi_{2}^{0}(\delta \xi_{3}^{0}+\delta \xi_{4}^{0})(u_{1i}^{*}v_{1j}^{*}+v_{1i}^{*}u_{1j}^{*})\nonumber \\
&+& g_{12}[(\phi_{1}^{0}\delta \xi_{4}^{0}+\phi_{2}^{0} \delta \xi_{2}^{0})(v_{1i}^{*}v_{2j}^{*})+(\phi_{1}^{0}\delta \xi_{3}^{0}+\phi_{2}^{0} \delta \xi_{1}^{0})(u_{1i}^{*}u_{2j}^{*})]\nonumber \\
&+& g_{12}[(\phi_{1}^{0}\delta \xi_{3}^{0}+\phi_{2}^{0} \delta \xi_{2}^{0})(v_{1i}^{*}u_{2j}^{*})+(\phi_{1}^{0}\delta \xi_{4}^{0}+\phi_{2}^{0} \delta \xi_{1}^{0})(u_{1i}^{*}v_{2j}^{*})]]
\end{eqnarray}

The total damping coefficient $\gamma$ is given by the imaginary part of the right-hand side of Eqn.(33). The damping coefficient $\gamma=\gamma_{L}+\gamma_{B}$ has two contributions, namely the Landau part ($\gamma_{L}$) and secondly the Baliaev part($\gamma_{B}$).

The expression for the Landau part is

\begin{equation}
\gamma_{L}=\pi \sum_{i,j}|A_{ij}|^{2} (f_{i}^{0}-f_{j}^{0}) \delta(\hbar \omega_{0}+\epsilon_{i}-\epsilon_{j})
\end{equation}

The Landau type damping arises when one quantum of oscillation $\hbar \omega_{0}$ being absorbed by a thermal excitation with energy $\epsilon_{i}$, which in turn produces another thermal excitation of energy $\epsilon_{j}$=$\epsilon_{i}+\hbar \omega_{0}$. In the one component case, this mechanism occurs in the same component. Interestingly, in the two component case, we find that the destruction and creation of the thermal excitations can also occur in two different components. From the expression of $A_{ij}$, one notices the following new possibilities due to the coupling $g_{12}$:

(1) One quantum of oscillation of the $m^{th}$ component being absorbed by a thermal excitation of the $n^{th}(n\neq m)$ component with the production of another thermal excitation in the $n^{th}$ component.
(2) one component of oscillation of the $m^{th}$ component being absorbed by a thermal excitation of the $m^{th}$ component with the production of another thermal excitation in the $n^{th}(n\neq m)$ component.
(3) one component of oscillation of the $m^{th}$ component being absorbed by a thermal excitation of the $n^{th}(n\neq m)$ component with the production of another thermal excitation in the $m^{th}$ component.

In the Baliaev type damping, a quantum of oscillation $\hbar \omega_{0}$ is absorbed and two excitations with energies $\epsilon_{i}+\epsilon_{j}=\hbar \omega_{0}$ are created.

The expression for the Baliaev part is

\begin{equation}
\gamma_{B}=\pi \sum_{i,j}|B_{ij}|^{2} (1+f_{i}^{0}+f_{j}^{0}) \delta(\hbar \omega_{0}-\epsilon_{i}-\epsilon_{j})
\end{equation}

Again from the expression for $B_{ij}$, we get the following new types  of Baliaev mechanisms due to the coupling $g_{12}$:

(1) Absorption of a condensate mode in the $m^{th}$ component is accompanied by creation of two excitations in the $n^{th}(n\neq m)$ component.

(2) Absorption of a condensate mode in the $m^{th}$ component is accompanied by creation of one excitation in the $m^{th}$ component and creation of the second excitation in the $n^{th}(n\neq m)$ component.

In addition, we find that the expressions for $\gamma_{L}$ and $\gamma_{B}$ which are proportional to $A_{ij}$ and $B_{ij}$ respectively, contain additional terms due to the inter-component coupling $g_{12}$. This clearly indicates enhanced damping in the two-component case compared to the single component case.
 We have thus shown that due to two-body inter-component interaction, coupling of the condensate modes of each component to the thermal excitations of the other component leads to new possibilities of Landau and Baliaev damping mechanism. The calculation of the Landau and Baliaev damping rates require the knowledge of the collective modes and the Bogoliubov amplitudes, which has recently been calculated \citep{56}.

\section{Conclusion}
We have studied the Landau and the Baliaev damping of the collective modes excited in a two component Bose-Einstein condensate. The two body inter-component interaction couples the collective mode of the system to the thermal fluctuations of both the components. This leads to various types of Landau and  Baliaev damping mechanism in which the creation or destruction of the elementary excitations can take place in the two separate components. We have also shown that coupling between the two components leads to enhanced damping.

\section{Acknowledgements}

A. Bhattacherjee acknowledge financial support from the Department of Science and Technology, New Delhi for financial assistance vide grant SR/S2/LOP-0034/2010.

\end{document}